\documentclass[prb,twocolumn,superscriptaddress]{revtex4-1}

\usepackage{amsmath}
\usepackage{amssymb}
\usepackage{xspace}
\usepackage{graphicx}
\usepackage{grffile}
\usepackage{nicefrac}
\usepackage{stmaryrd}
\usepackage{color}

\newcommand{\bw}[1]{\raisebox{1.5ex}[-1.5ex]{#1}}
\begin{document}

\title{Quantum-Many-Body Intermetallics: Phase Stability of Fe$_3$Al and Small-Gap
Formation in Fe$_2$VAl}

\author{Oleg Kristanovski}
\affiliation{I. Institut f{\"u}r Theoretische Physik, Universit{\"a}t Hamburg, 
D-20355 Hamburg, Germany}
\author{Raphael Richter}
\affiliation{Institut f\"ur Technische Thermodynamik, Deutsches Zentrum f\"ur Luft-
und Raumfahrt, D-70569 Stuttgart}
\author{Igor Krivenko}
\affiliation{I. Institut f{\"u}r Theoretische Physik, Universit{\"a}t Hamburg, 
D-20355 Hamburg, Germany}
\author{Alexander I. Lichtenstein}
\affiliation{I. Institut f{\"u}r Theoretische Physik, Universit{\"a}t Hamburg, 
D-20355 Hamburg, Germany}
\author{Frank Lechermann}
\affiliation{I. Institut f{\"u}r Theoretische Physik, Universit{\"a}t Hamburg, 
D-20355 Hamburg, Germany}
\affiliation{Institut f{\"u}r Keramische Hochleistungswerkstoffe, Technische Universit\"at
Hamburg-Harburg, D-21073 Hamburg, Germany}


\begin{abstract}
Various intermetallic compounds harbor subtle electronic correlation effects. To elucidate
this fact for the Fe-Al system, we perform a realistic many-body investigation based on 
the combination of density functional theory with dynamical mean-field theory in a charge
self-consistent manner. A better characterization and understanding of the phase stability of 
bcc-based D0$_3$-Fe$_3$Al through an improved description of the correlated charge density 
and the magnetic energy is achieved.
Upon replacement of one Fe sublattice by V, the Heusler compound Fe$_2$VAl is realized, 
known to display bad-metal behavior and increased specific heat. We here document a
charge-gap opening at low temperatures in line with previous experimental work. The gap
structure does not match conventional band theory and is reminiscent of (pseudo)gap
charateristics in correlated oxides.
\end{abstract}

\maketitle

\section{Introduction}
The Fe-Al system is well-known for its intricate phase diagram, displaying a 
complex interplay between metallicity, magnetism and structure. 
Stoichiometric FeAl poses a longstanding problem regarding its magnetic ground state. While 
experimentally B2-FeAl is characterized as a Curie-Weiss paramagnet~\cite{miy68} with no 
detectable ordered moment, conflicting results exist in 
theory.~\cite{min86,moh01,pet03,gal15} On the Al-rich
side, the low-symmetry structures FeAl$_2$ and Fe$_2$Al$_5$ exhibit spin-glass physics
at low temperature.~\cite{lue01,jag11} On the iron-rich side at Fe$_3$Al composition 
a bcc-based D0$_3$ crystal structure is stable with well-defined ferromagnetic (fm) order up 
to $T_{\rm c}=$713\,K.~\cite{fal36} Further increase of the Fe content transforms the system 
into the doped bcc-Fe (or $\alpha$-) phase, also with fm order below a Curie temperature of 
1043\,K for pure iron.
Albeit unambiguous in nature, both $\alpha$-Fe and D0$_3$-Fe$_3$Al appear difficult to be 
described within conventional density functional theory (DFT).~\cite{wan85,bag89,lec02} 
The generalized-gradient approximation (GGA) for the exchange-correlation energy is
 necessary to detect the fm-bcc-Fe ground state.~\cite{bag89}
Intriguingly, the fm-D0$_3$ compound is only stable within the local-density approximation 
(LDA), while GGA favors the fcc-based L1$_2$ structure in the ferromagnetic 
state.~\cite{lec02}

This lack of coherent theoretical description of the Fe-rich side of Fe-Al in standard 
Kohn-Sham DFT asks for extended approaches. The inclusion of static electronic correlation 
effects via the DFT+Hubbard $U$ method may cope with part of the subtle energetics for a 
reasonable choice of the local Coulomb-interaction parameters.~\cite{lec04} But that scheme 
is in principle not well defined for correlated itinerant systems and in addition usually 
needs to enforce magnetic order to deliver proper results. True paramagnetic (pm) states 
based on fluctuating local moments are neither accessible in conventional DFT nor in DFT+U,
which either describes nonmagnetic (nm) or magnetically ordered compounds. Within the 
so-called disordered-local-moment (DLM) method~\cite{gyo85,abr15} there is the chance to 
account for a DFT-based orientational mean-field effect of pm-like spins. Yet quantum 
fluctuations as well as general finite-temperature fluctuations of e.g. the size of the 
local moments are still missing.

A further facet of the intriguing correlated electronic structure in iron aluminides is 
revealed when replacing one Fe sublattice in D0$_3$-Fe$_3$Al by vanadium. This transforms 
the intermetallic crystal to the Heusler L2$_1$ compound Fe$_2$VAl. The ordered alloy is 
paramagnetic down to lowest temperatures and displays bad-metal behavior in 
transport.~\cite{nis97} It is still a matter of debate if Fe$_2$VAl is a small-gap 
($\sim 0.1-0.3$\,eV) semiconductor or a semimetal.~\cite{lue98,oka00} 
Reminiscent of $f$-electron systems like SmB$_6$ with Kondo-insulating 
characteristics,~\cite{all79} heavy-fermion physics was originally associated with 
this $3d$-electron system.~\cite{nis97,lue98} Though magnetic defects later explained
a sizable part of the large specific heat at low temperature, the overall mass
enhancement remains substantial.~\cite{lue99}
Promising thermoelectric potential due to an enhanced thermopower is associated with
Fe$_2$VAl-based materials.~\cite{nis06,mik09} Again, the theoretical 
first-principles assessment is difficult, since e.g. there are substantial differences 
concerning the existence of a charge gap $\Delta$ and its eventual size. Conventional DFT 
based on LDA/GGA classifies Fe$_2$VAl as semimetallic,~\cite{sin98,weh98} 
whereas the use of hybrid functionals renders the system semiconducting with a band gap of 
$\Delta_{\rm g}=0.34$\,eV.~\cite{bil11} A gap of $\Delta_{\rm g}=0.55$\,eV is revealed
from DFT+U calculations.~\cite{do11}

In this work a first-principles many-body approach is employed to consider the effects of
quantum fluctuations and finite temperature on the electronic structure of Fe$_3$Al and 
Fe$_2$VAl beyond conventional DFT(+U). The state-of-the-art combination of density 
functional theory with dynamical mean-field theory (DMFT) reveals important modifications 
of the correlated electronic structure. We show that the subtle electronic states  
rely on many-electron quantum processes, with important consequences for the phase stability 
and tendencies concerning gap formation. This paves the road towards a coherent 
modeling and understanding of Fe-Al and signals the general importance of advanced 
theoretical schemes for intermetallic systems.

\section{Crystal Structures}
\begin{figure}[b]
\includegraphics*[width=8.5cm]{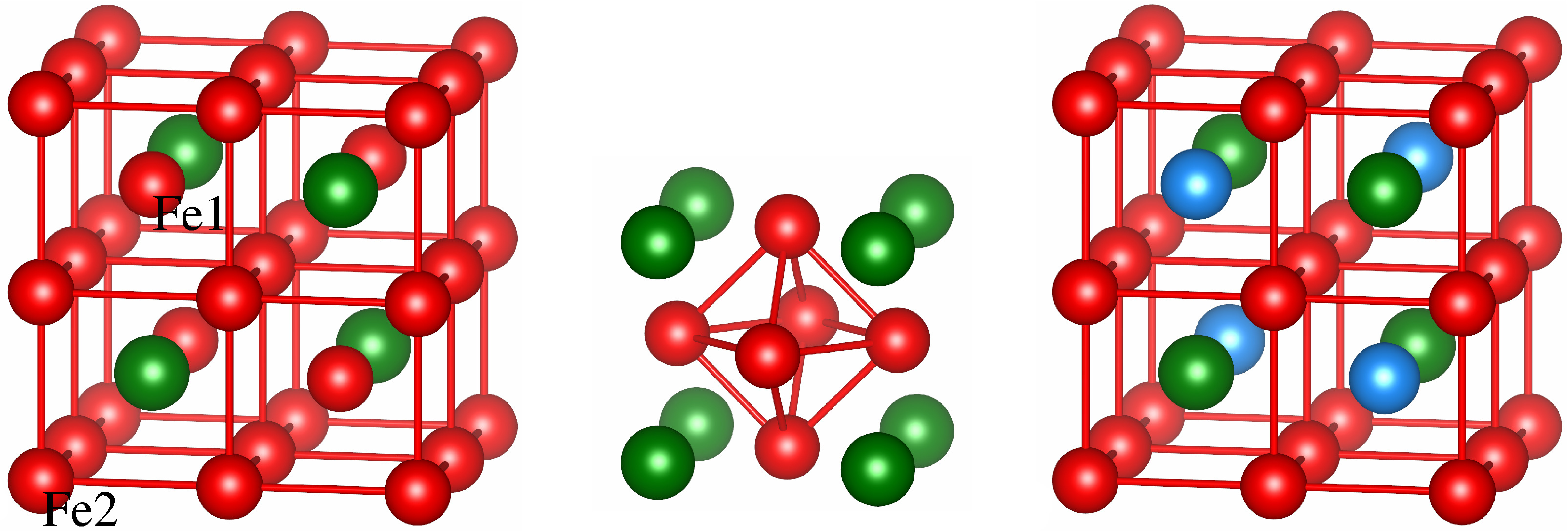}
\caption{(color online) Relevant crystal structures. From left to right:
D0$_3$-Fe$_3$Al, L1$_2$-Fe$_3$Al and L2$_1$-Fe$_2$VAl. Fe (red/lightgrey), Al (green/dark)
and V (blue/grey).}\label{fig1:struc}
\end{figure}
The crystal structures relevant for this work are displayed in Fig.~\ref{fig1:struc}.
With bcc-Fe and fcc-Al as end members, the cubic lattice system also accounts for the 
dominant ordered phases in Fe-Al. Starting with B2-FeAl at stoichiometry, the bcc 
lattice is the common host for the stable solid phases in the Fe-rich part. Though the 
D0$_3$ structure is stable in the Fe$_3$Al phase regime, the fcc-based L1$_2$ 
structure appears as a relevant competitor. The D0$_3$ unit cell consists of three Fe
and one Al site, whereby the Fe basis atoms are grouped in two symmetry shells. One Fe site 
belongs to the Fe1 sublattice and two Fe sites to the Fe2 sublattice. As a bcc structure,
each Fe site has 8 nearest-neigbor (NN) sites. Whereas the Fe2 atoms
have mixed Fe/Al nearest neighborhood, the Fe1 atom has only Fe nearest neighbors. The
experimental lattice constant of fully-ordered Fe$_3$Al reads $a=5.473$\, a.u.. 

The L1$_2$ structure also consists of three Fe and one Al atom in the primitive unit cell, 
but all Fe sites are equivalent by symmetry. The 12-atom NN shell of these Fe sites
is composed again of both, Fe and Al sites.

Finally, in the Heusler L2$_1$-Fe$_2$VAl compund the Fe1 sublattice of the D0$_3$ structure 
is fully replaced by V atoms. The measured lattice constant amounts to 
$a'=5.442$\,a.u.~\cite{nis97,lue98}

Note that troughout this work we investigate the stiochiometric compounds, 
i.e. defect physics and effects of chemical disorder are not treated.

\section{Computational Framework}
The charge self-consistent DFT+DMFT methodology~\cite{sav01,pou07,gri12} is put into 
practise, utilizing a mixed-basis 
pseudopotential approach~\cite{lou79,mbpp_code} for the DFT part and the continuous-time 
quantum-Monte-Carlo scheme~\cite{rub05,wer06} from the TRIQS package~\cite{par15,set16}
for the DMFT impurity treatment. Exchange-correlation in the Kohn-Sham cycle is handeled by
the GGA functional of Perdew-Burke-Ernzerhof (PBE)~\cite{per96} form. 
\begin{figure}[t]
\includegraphics*[width=8.5cm]{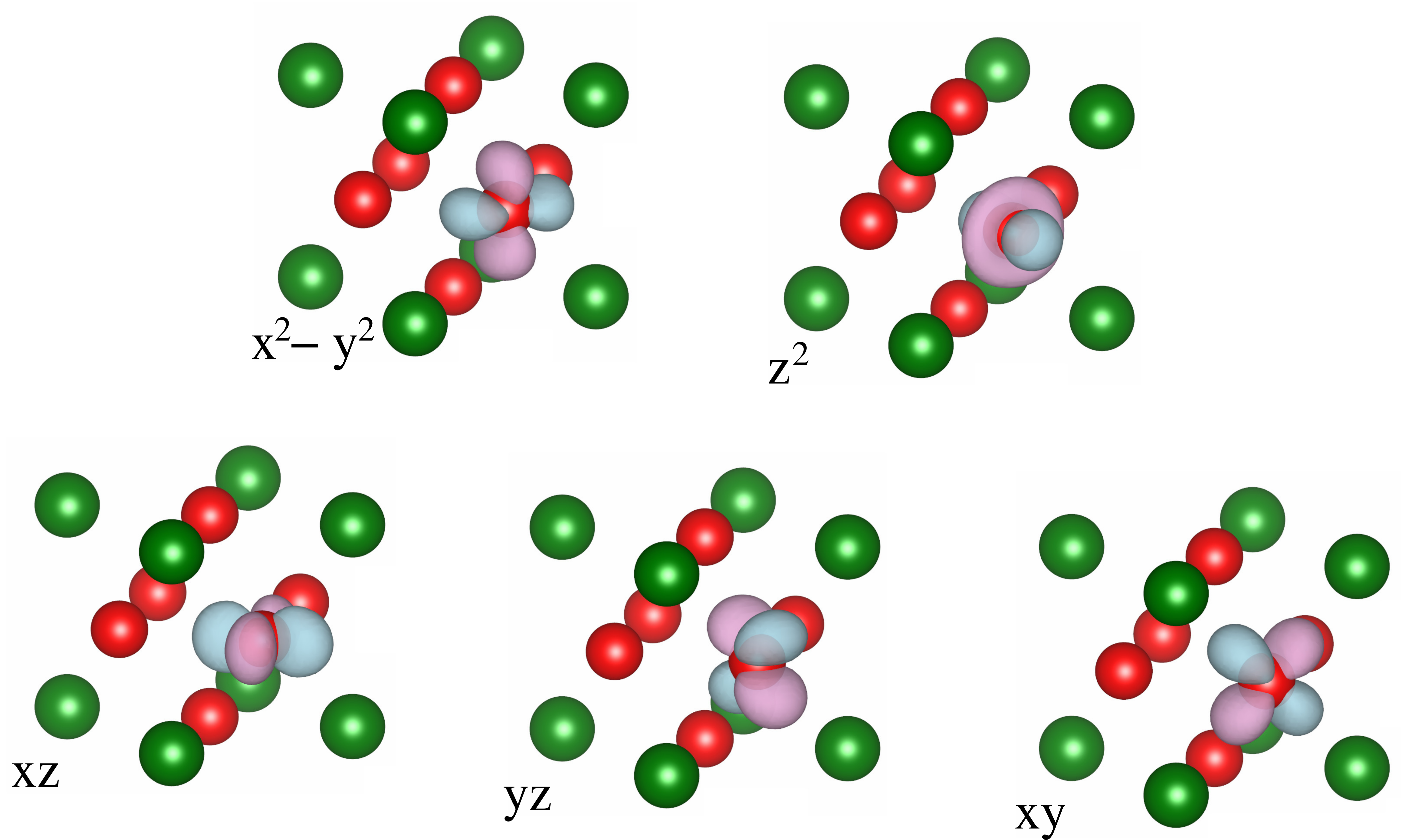}\\
\caption{(color online) Projected local Fe($3d$) orbitals in L1$_2$-Fe$_3$Al.
On-site level energies: $\varepsilon_{\alpha}=\{-624,-699,-843,-843,-998\}\,$meV for
the effective orbitals $\alpha=\{x^2$$-$$y^2,z^2,xz,yz,xy\}$.}\label{fig2:l12orbs}
\end{figure}

The correlated subspace where quantum fluctuations are explicitly accounted for is associated 
with the transition-metal sites of Fe and V kind. Projected-local 
orbitals~\cite{ama08,ani05,aic09,hau10,kar11} of $3d$ character are used to extract 
Wannier-like states based on 22 Kohn-Sham bands, stemming from Fe/V($3d\,4s$) and 
Al($3s\,3p$) orbitals. Each transition-metal site represents a DMFT impurity problem, which 
due to symmetry amounts to two such ones in D0$_3$-Fe$_3$Al and Fe$_2$VAl, while only one 
symmetry-inequivalent transition-metal site is hosted in L1$_2$-Fe$_3$Al. A multi-orbital
Hubbard Hamiltonian of Slater-Kanamori form, parametrized by the Hubbard $U$ and the Hund's 
exchange $J_{\rm H}$, is applied to the respective full 5-orbital $3d$-manifold. We 
overtook the values $U=3.36$\,eV and $J_{\rm H}=0.71$\,eV for the local Coulomb 
interactions from Ref.~\onlinecite{gal15}, where those are computed for 
B2-FeAl using the constrained random-phase approximation. 
A double-counting correction of the fully-localized form is used in this work. If 
not stated otherwise, the temperature within the DMFT part is set to $T=387\,$K, i.e. 
$\beta=1/T=30\,$eV$^{-1}$. The analytical continuation of the Green's functions on the 
Matsubara axis $i\omega$ is performed via the maximum-entropy method.

We mainly focus in our DMFT calculations on phases without broken spin symmetry, i.e. 
paramagnetic states. Albeit D0$_3$-Fe$_3$Al is ferromagnetic at ambient temperatures, the 
explicit magnetic {\sl ordering} energy, as will be shown below, is not of vital 
importance for our investigation and its conclusions.
 
\section{Results}

\subsection{Fe$_3$Al}

\subsubsection{Electronic states}
\begin{figure}[b]
(a)\hspace*{-0.4cm}\includegraphics*[width=8.65cm]{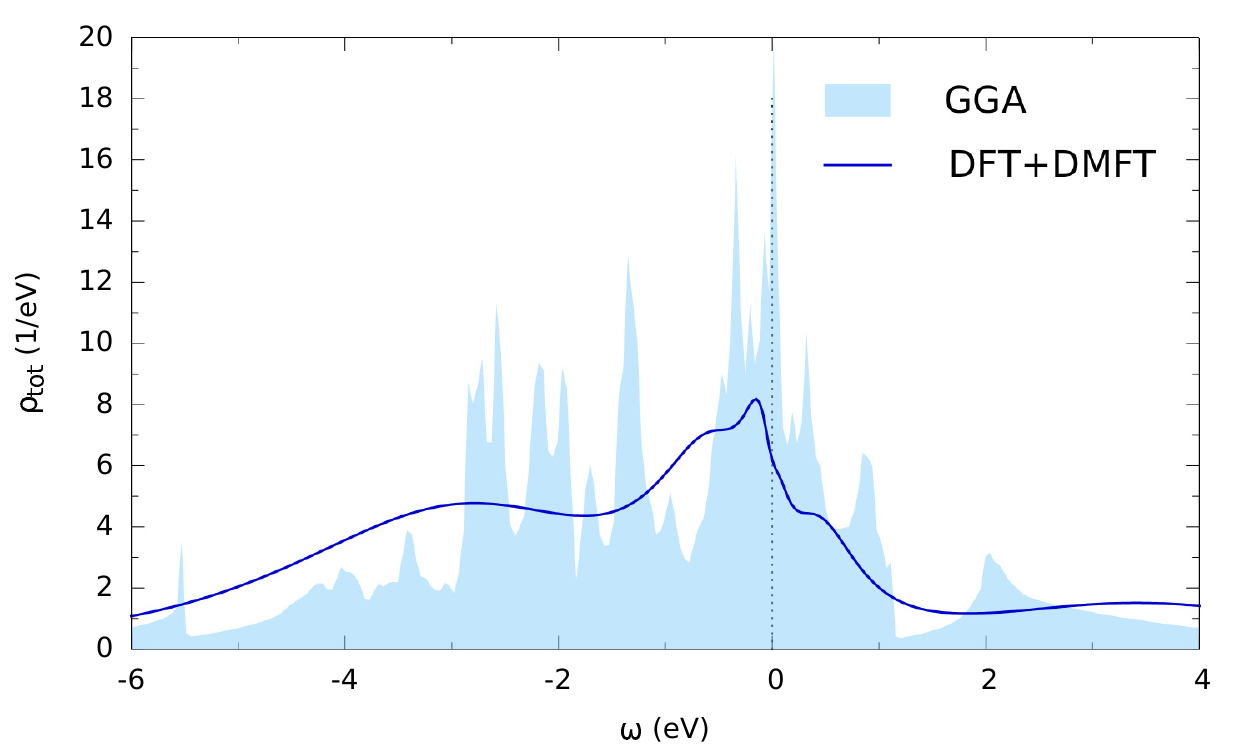}\\
(b)\hspace*{-0.4cm}\includegraphics*[width=8.65cm]{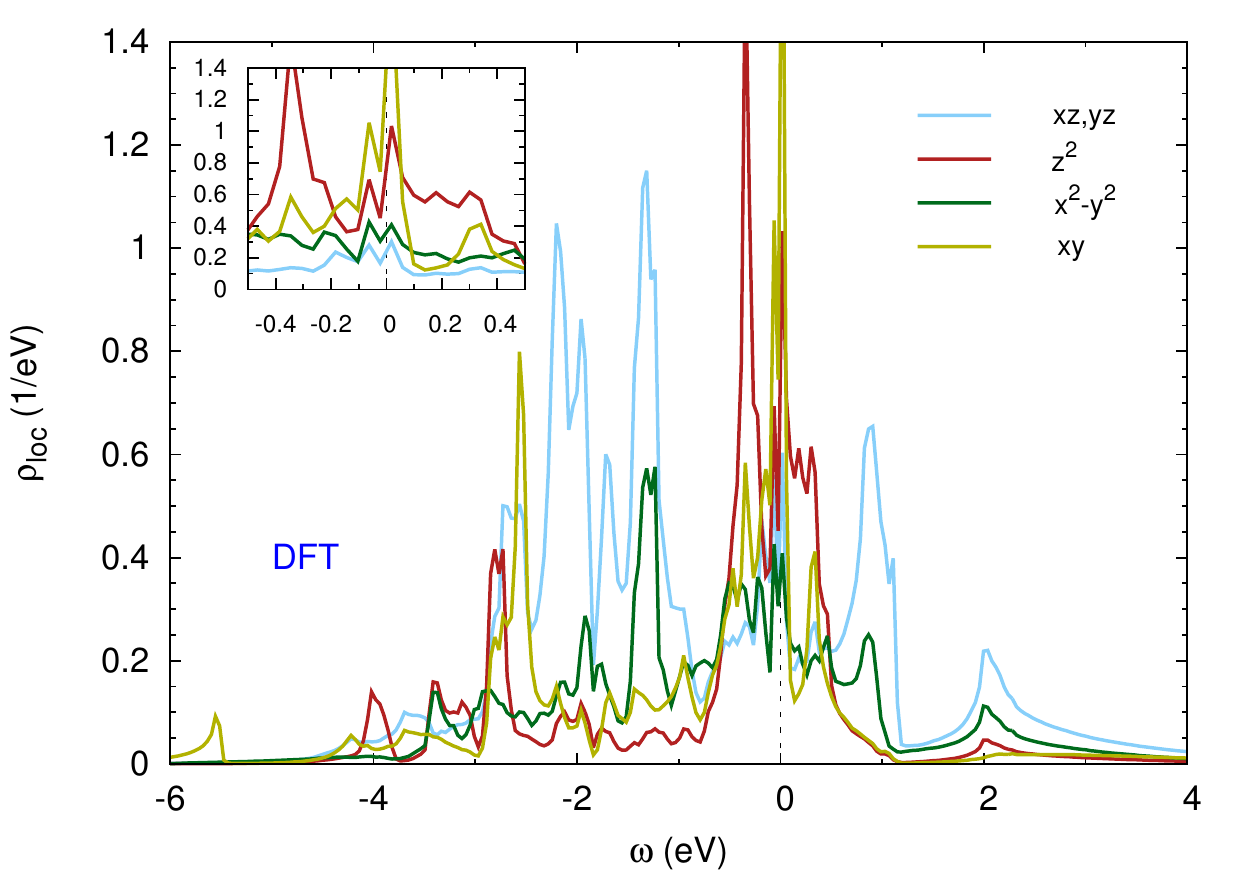}\\
(c)\hspace*{-0.4cm}\includegraphics*[width=8.65cm]{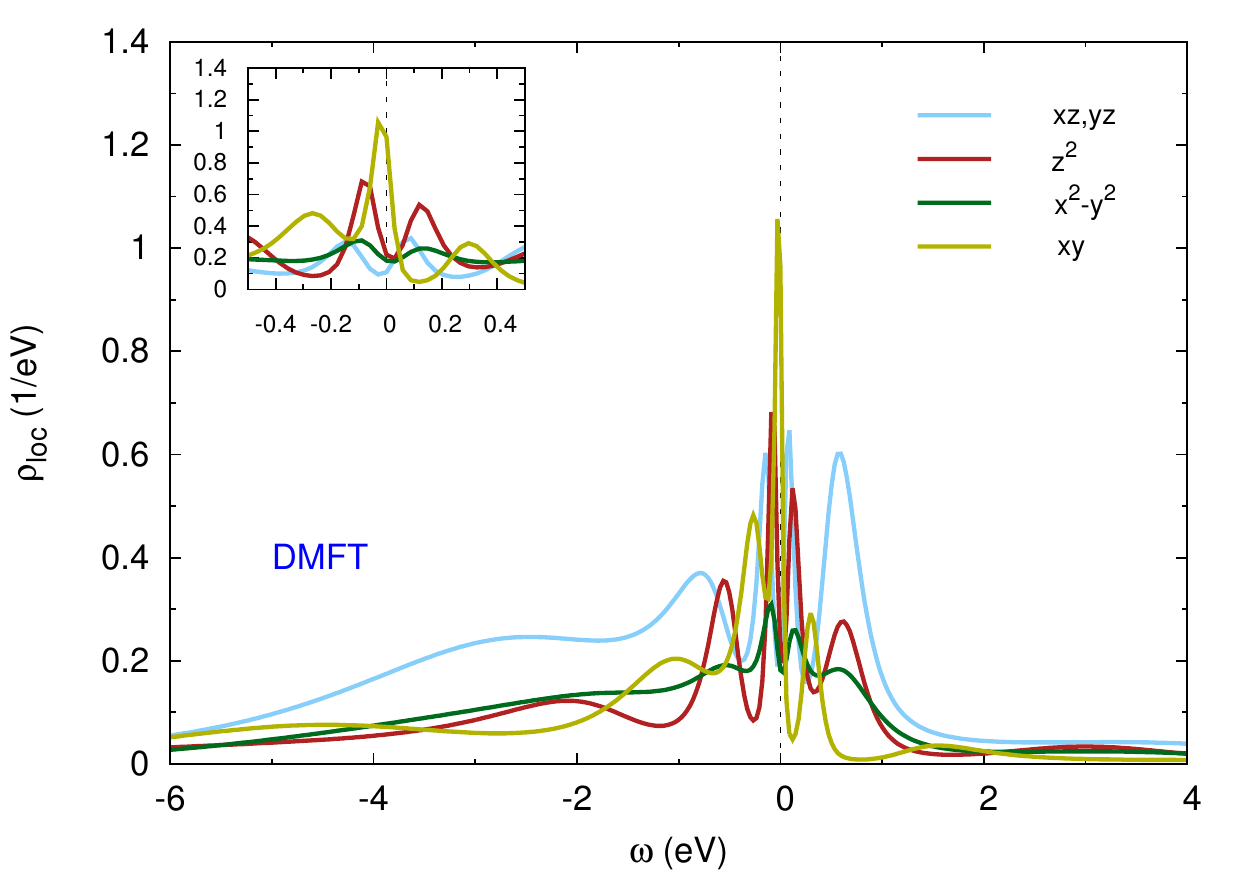}
\caption{(color online) Spectral functions of L1$_2$-Fe$_3$Al.
(a) Total, (b) local Fe from GGA and (c) local Fe from DFT+DMFT.
Insets in (b,c) are low-energy blow ups.}\label{fig3:l12spec}
\end{figure}
\begin{figure}[b]
(a)\hspace*{-0.4cm}\includegraphics*[width=8.65cm]{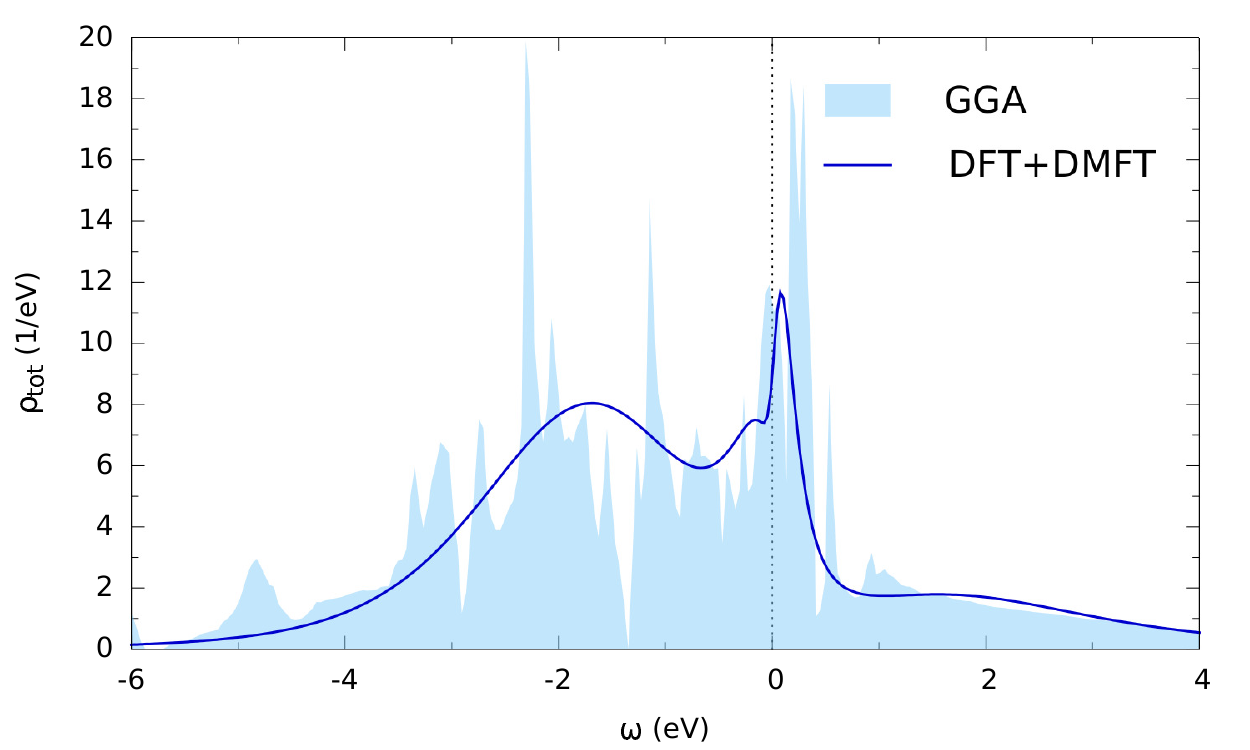}\\
(b)\hspace*{-0.4cm}\includegraphics*[width=8.65cm]{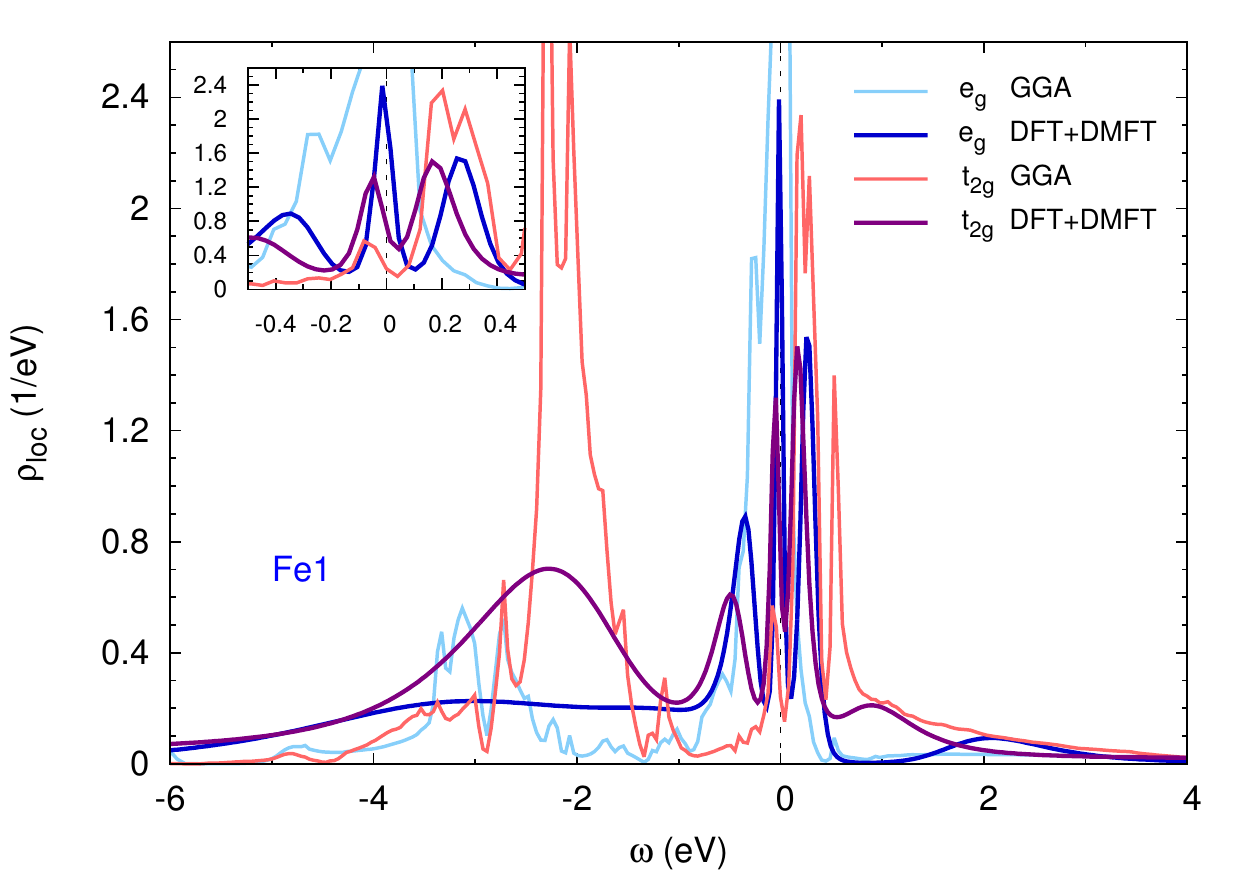}\\
(c)\hspace*{-0.4cm}\includegraphics*[width=8.65cm]{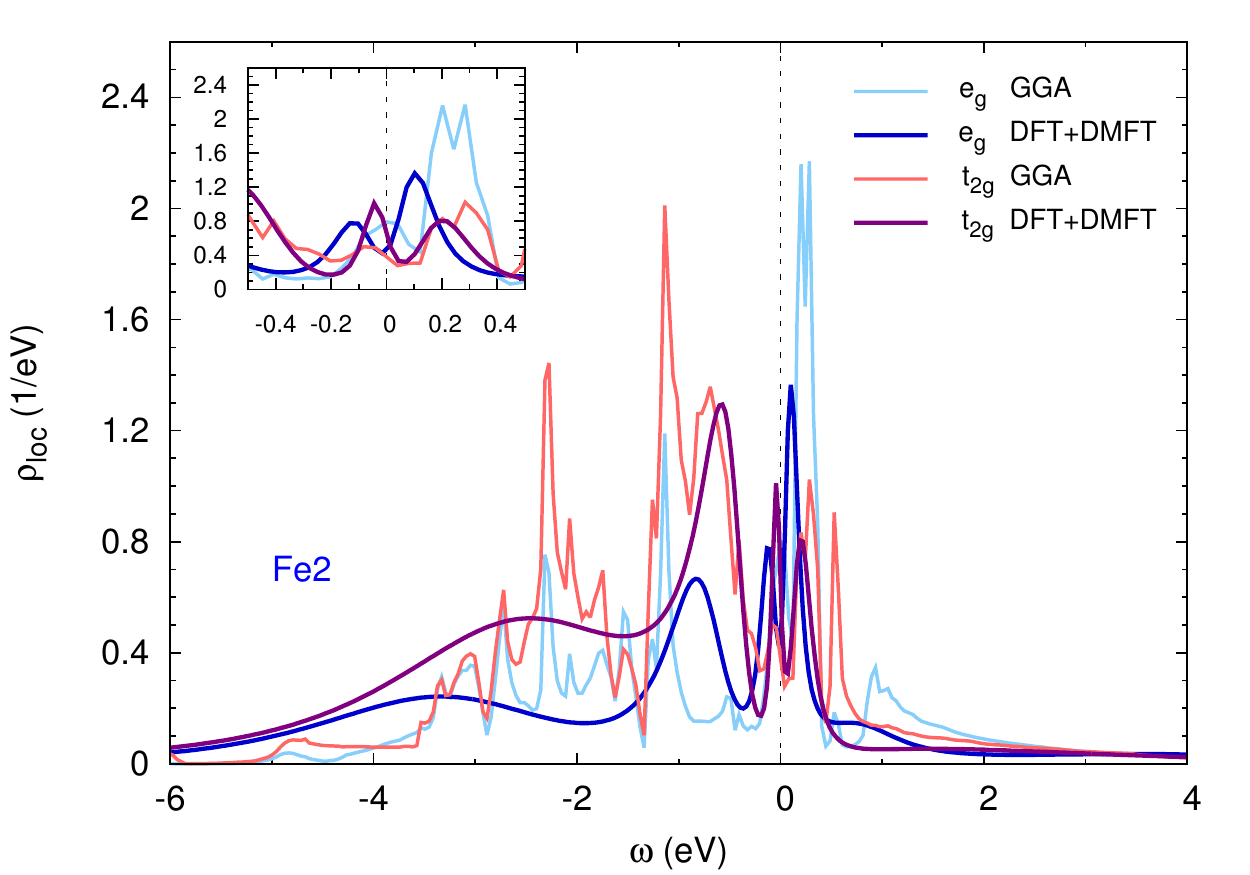}\\
\caption{(color online) Spectral functions of D0$_3$-Fe$_3$Al.
(a) Total, (b) local Fe1 and (c) local Fe2. Insets in (b,c) are low-energy blow
ups.}\label{fig4:d03spec}
\end{figure}
We first focus in some detail on the electronic states in Fe$_3$Al. Let us start with the
fcc-based L1$_2$ structure, having only one transition-metal sublattice. The close-packed
lattice type is an important one in intermetallic systems, e.g. the ordered phases Cu$_3$Au
and Ni$_3$Al condense in the L1$_2$ structure. Because of the cubic symmetry, here the local
Fe($3d$) states in principle split twofold into $e_g$ and $t_{2g}$ states. However due to
the ordering pattern, not all $e_g$/$t_{2g}$ sublevels may still be degenerate. This is
illustrated in Fig.~\ref{fig2:l12orbs}, where the obtained Fe($3d$) projected local orbitals
are plotted as isosurfaces. The $e_g$ manifold consisting of $\{x^2$$-$$y^2,z^2\}$ is defined
by the orbital lobes pointing towards the next-nearest neighboring (NNN) Fe sites. Since both
pointing directions are anisotropic in terms of the respective nearest-neighbor sites,
the two $e_g$ are non-degenerate. The $t_{2g}$ manifold consisting of $\{xz,yz,xy\}$ are
defined by the orbital lobes pointing to the NN sites. For $xz,yz$ the associated NN sites
are exclusively of Fe kind, therefore both orbitals are degenerate. Yet in the case of $xy$
the associated NN shell consists exclusively of Al sites, thus this $t_{2g}$ orbital has
a different, in fact the lowest effective, crystal-field level.
\begin{table}[b]
\begin{ruledtabular}
\begin{tabular}{l|lcccccr}
             &          & $e_{g}$   &       & $t_{2g}$ &    &      &  total\\
             &          & $x^2-y^2$ & $z^2$ & $xz$  & $yz$ & $xy$ &  \\ \hline
             &          & 1.38      & 1.34  & 1.43  & 1.43 & 1.58 &\hspace*{0.5cm}7.16\\
\bw{L1$_2$}  & \bw{Fe}  & 1.55      & 1.56  & 1.54  & 1.54 & 1.79 &\hspace*{0.5cm}7.98\\ \hline
             &          & 1.45      & 1.45  & 1.31  & 1.31 & 1.31 &\hspace*{0.5cm}6.83\\
             & \bw{Fe1} & 1.47      & 1.47  & 1.54  & 1.54 & 1.54 &\hspace*{0.5cm}7.56\\[0.1cm]
\bw{D0$_3$}  &          & 1.20      & 1.20  & 1.59  & 1.59 & 1.59 &\hspace*{0.5cm}7.17\\
             & \bw{Fe2} & 1.45      & 1.45  & 1.70  & 1.70 & 1.70 &\hspace*{0.5cm}8.00\\
\end{tabular}
\end{ruledtabular}
\caption{Projected-local-orbital occupations in Fe$_3$Al. First lines are
GGA, second lines DFT+DMFT results, respectively.}\label{tab1:occ}
\end{table}

Figure~\ref{fig3:l12spec} compares the integrated spectral functions
$\rho(\omega)=\sum_{\bf k}A({\bf k},\omega)$ of L1$_2$-Fe$_3$Al within DFT(GGA)
and DFT+DMFT. From the broadly itinerant band structure, an effective
relevant bandwidth of about 7 eV (ranging from $-6$ eV to 1 eV) may be read off.
Seemingly the full Fe($3d$) manifold is cruicial to understand the electronic structure
in the bonding part and at low energy, since the hybridization between Fe and Al is rather
strong in a wide energy range. Close to the Fermi level, the $z^2$ and $xy$ effective
orbital are most dominant in GGA, while e.g. the $xz/yz$ part displays a
bonding-antibonding signature.

For the $xy$ state with deepest crystal-field level and
broad dispersion, the orbital filling is also largest (see Tab.~\ref{tab1:occ}).
The total local Fe electron count is slightly above seven within GGA. A further
strengthening of the $xy$ dominance at low-energy occurs in the DFT+DMFT treatment.
While the filling of all effective Fe orbitals increases with correlations, also here
the occupation of the $xy$ state is enhanced largest by relative means. Overall, a
substantial increase in the total effective Fe($3d$) filling close to eight electrons
takes place. Note that the site-filling differences between GGA and DFT+DMFT are also
due to the respective effective-orbital definitions, as usual in determining local
occupations in crystalline solids. First, the projected-local orbitals in both calculational
schemes are not identical (only the projecting functions are), since via the charge
self-consistent loop the Kohn-Sham part (i.e. the bands used for the projection) changes.
Second, the resulting orbitals are of Wannier kind, i.e. their spread is substantial and
not localized on the site centre within a small spherical radius.
\begin{figure}[t]
\hspace*{-0.2cm}(a)\includegraphics*[width=3.9cm]{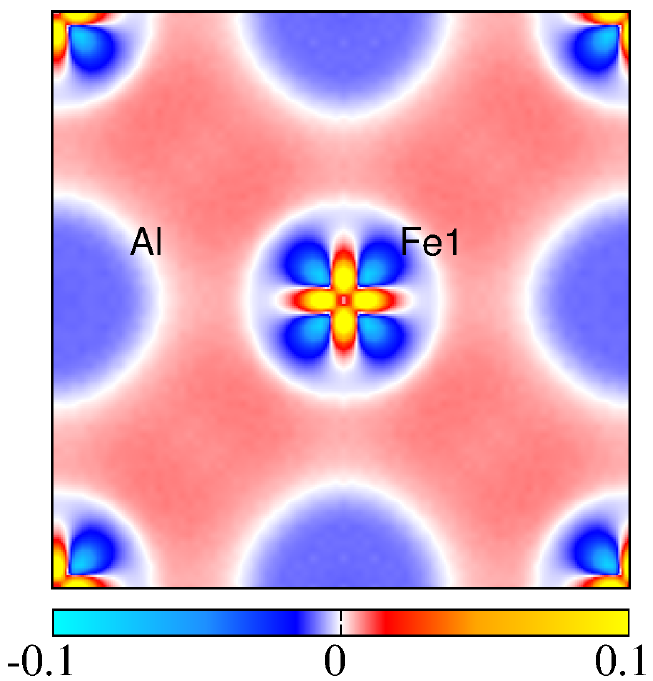}\hspace*{0.1cm}
\includegraphics*[width=3.9cm]{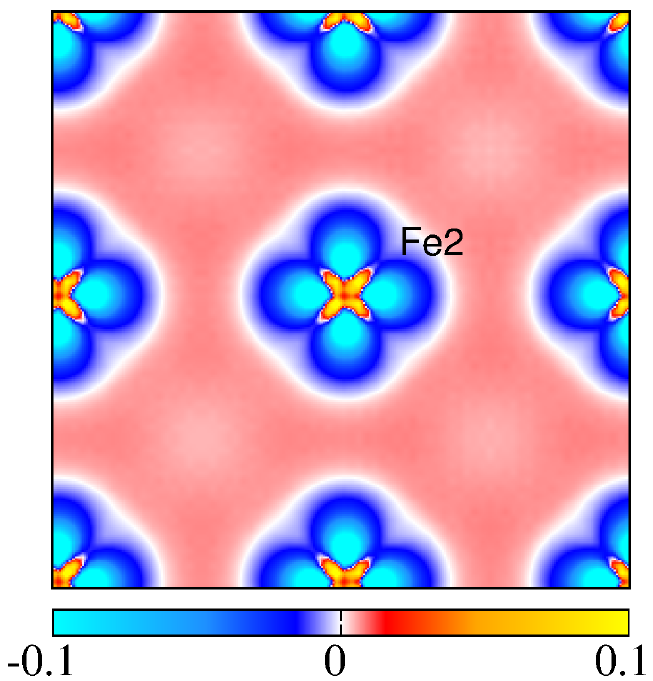}
\hspace*{-0.2cm}(b)\includegraphics*[width=3.9cm]{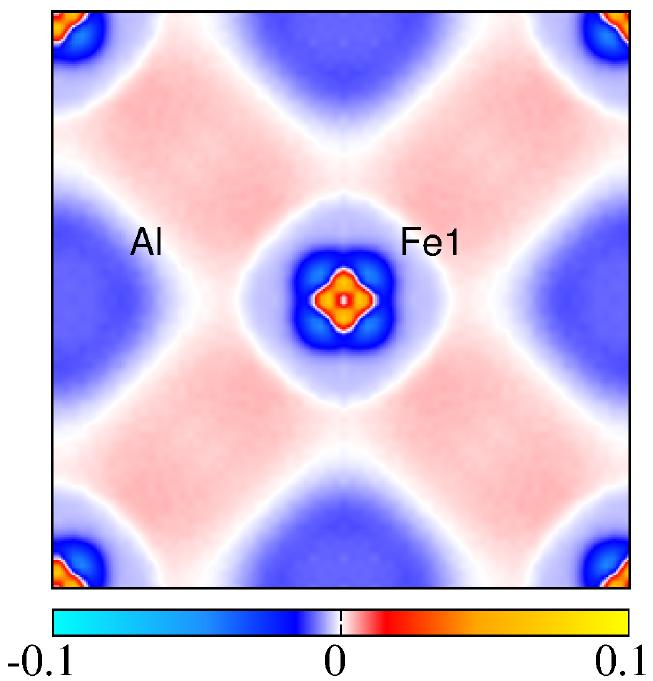}\hspace*{0.1cm}
\includegraphics*[width=3.9cm]{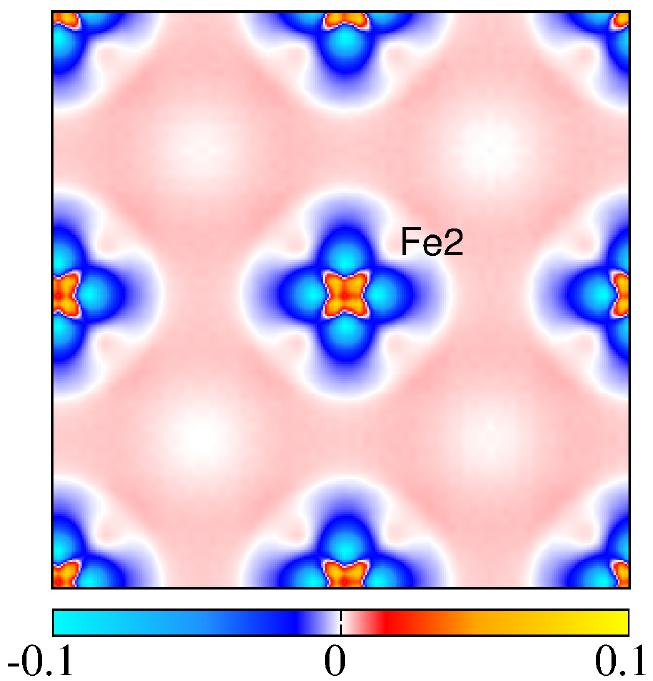}
\hspace*{-0.2cm}(c)\includegraphics*[width=3.9cm]{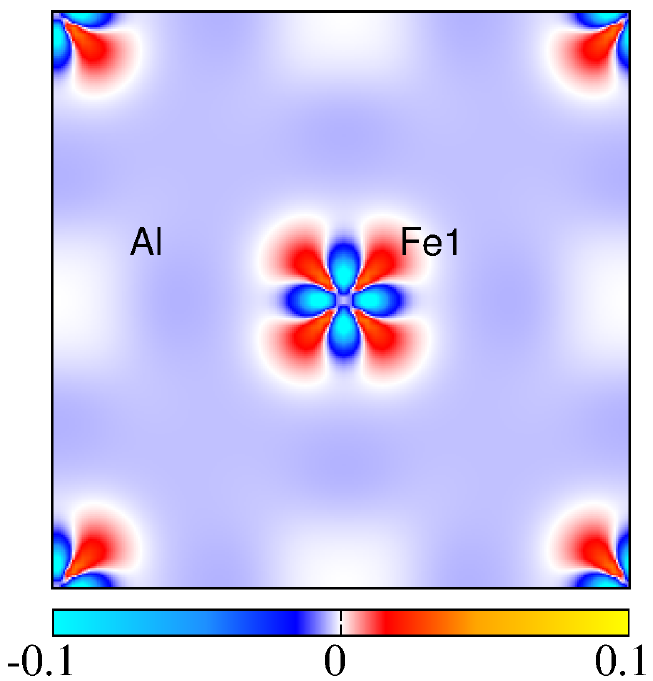}\hspace*{0.1cm}
\includegraphics*[width=3.9cm]{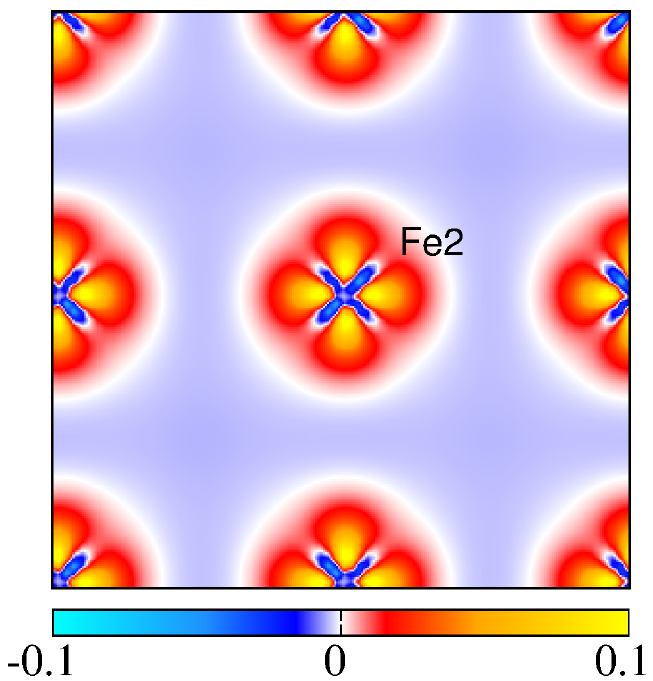}
\caption{(color online) Relevant charge-density plots in D0$_3$-Fe$_3$Al around
Fe1 (left) and Fe2 (right) with $c$-axis perpendicular to the plotting plane.
(a) GGA bonding charge density (see text). (b) DFT+DMFT bonding charge density.
(c) Charge difference $n_{\rm DFT+DMFT}-n_{\rm GGA}$.}\label{fig5:charge}
\end{figure}

Still correlations may enhance the electron localization on the Fe sites.
The correlation strength can be estimated from the quasiparticle (QP) weight 
$Z\sim 1/m_{\rm eff}$, computed from the electronic self-energy on the Matusbara axis as
\begin{equation}
Z=\left(1-\left.\frac{\partial\,{\rm Im}\Sigma(i\omega)}{\partial\,\omega}
\right|_{\omega\rightarrow0^+}\right)^{-1}\;.
\end{equation} 
There is no strong orbital variation of the QP weight in the L1$_2$ structure and 
it amounts to a moderate value of $Z\sim 0.7$.

Though the D0$_3$ structure consists of two different Fe sublattices, the 
conventional internal degeneracies of the $e_g$ and $t_{2g}$ subshells of Fe($3d$) 
are here fulfilled. This is due to the fact that the NN environments are either of pure 
Fe kind or of equally mixed-isotropic Fe,Al kind. As in fcc-based L1$_2$, the $e_g$ orbitals 
point again towards NN and NNN sites. However, since bcc-based D0$_3$ is 
not close packed, the $t_{2g}$ orbitals point inbetween the NN and NNN, i.e. 
towards the 3rd-nearest neighbor sites.

The total integrated spectral function of D0$_3$-Fe$_3$Al is similar to the one of 
L1$_2$-Fe$_3$Al (see Fig.~\ref{fig4:d03spec}a), but with a more pronounced quasiparticle
peak at low energy. The effective relevant bandwidth seems also smaller by about 1 eV 
in extent. On the local level, the Fe1 spectrum exhibts stronger $e_g$-$t_{2g}$
discrimination than the Fe2 spectrum. This speaks for a more subtle orbital/directional
electronic structure around Fe1, whereas Fe2 with its 'washed-out' orbital signature looks
similar to Fe in the L1$_2$ structure. A strong GGA favoring of $e_g$ character
at low energy in the case of Fe1 is weakened in DFT+DMFT, i.e. with correlations there
are orbital-balancing tendencies at the Fermi level. 

From the electron count, the
Fe1($t_{2g}$) states become strongly correlation-enhanced, while on the other hand
the Fe2($e_g$) electrons benefit from a local Coulomb interaction (cf. Tab.~\ref{tab1:occ}). 
In principle, localizing D0$_3$ electrons in effective $t_{2g}$ orbitals is understandable
from a charge-repulsion argument due to the orbital direction. Because of the stronger-hybridized
environment on Fe2 imposed by nearby Al, there that single-site argument is not easily
applicable. Note that the effective $e_g$ filling is levelled out
in DFT+DMFT between Fe1 and Fe2.
Figure~\ref{fig5:charge} underlines the present findings by inspecting the intra- and inter-site
charge transfers. The bonding charge density $n^{\rm bond}\equiv n^{\rm crystal}-n^{\rm atom}$ 
with many-body effects shows furthermore charge depletion in the interstitial bonding region 
compared to the GGA result.
In total, also the Fe sites in the D0$_3$ structure gain $3d$ electrons upon the impact
of local Coulomb interactions. While as expected the Fe2 site has a similar filling
as the Fe site in L1$_2$, the Fe1 site has a lower electron count by roughly half an
electron. Note that the absence of significant Fe filling differences with correlations
in the recent work by Galler {\sl et al.}~\cite{gal15} for B2-FeAl might be 
explained by the fact that no charge self-consistent framework was utilized in that study.

Concerning the correlation strength, though the Fe1 site and in general the $e_g$ orbital
character has a somewhat lower QP weight, there is neither striking difference between the 
two Fe sublattices, nor between the $e_g$/$t_{2g}$ character. In numbers, an average value 
of $Z\sim 0.8$ is slightly higher than for the L1$_2$ structure, marking somewhat weaker
correlation effects.

The Fe$_3$Al compound is not a particularly strongly correlated material, since the 
ratio of the local Coulomb interaction and the bandwidth is well below unity. In addition, 
the local Fe occupation ranging between seven and eight electrons is already above the 
optimal Hund's physics scenario~\cite{wer08,hau09,med11} of about $5\pm1$ electrons (where e.g. 
iron pnictides reside). Still correlation effects are effective in modifiying the charge 
density and the low-energy character, having impact on bonding properties as well as on charge 
and spin response.

\subsubsection{Energetics}
\begin{figure}[b]
\centering
\includegraphics*[width=8.5cm]{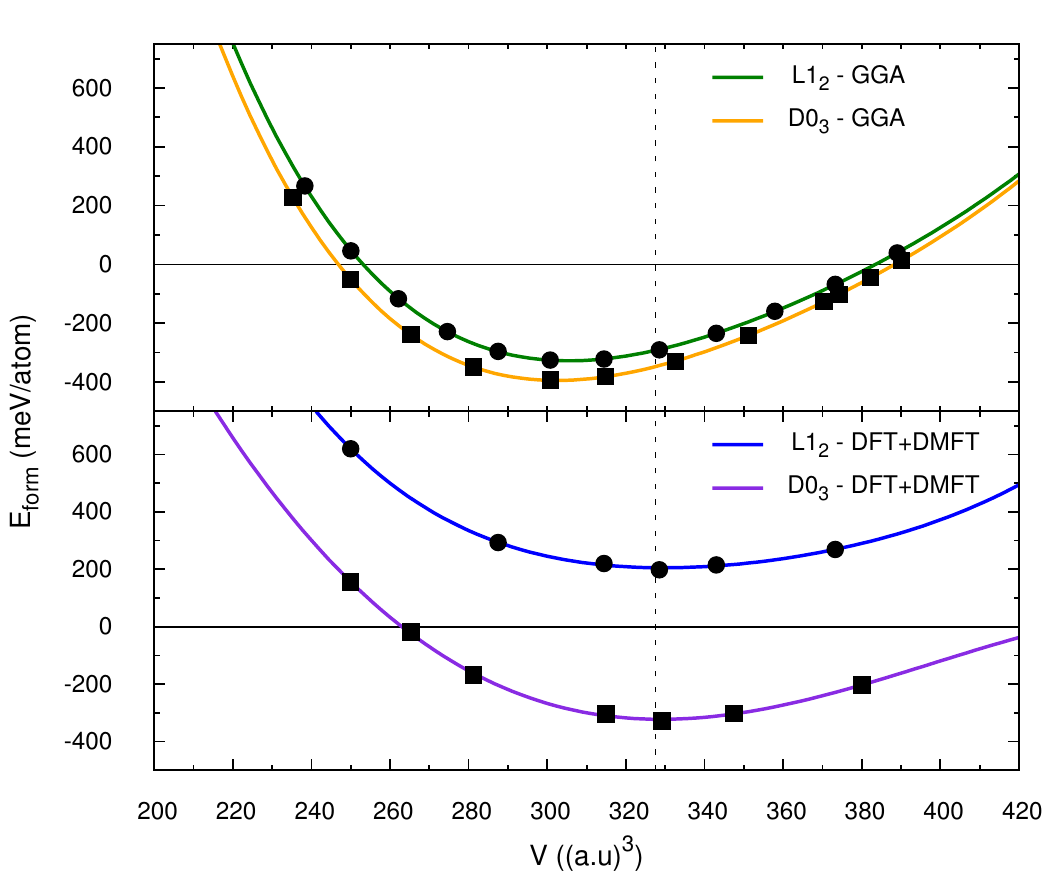}
\caption{(color online) Comparison of Fe$_3$Al formation-energy curves for 
the D0$_3$ and the L1$_2$ structure within nm-GGA and pm-DFT+DMFT. Dashed line marks the 
experimental equilibrium volume.}\label{fig2:toten}
\end{figure}

We turn now to the structural phase competition between D0$_3$ and L1$_2$, by comparing the 
formation energy $E_{\rm form}$ per atom with respect to the volume $V$, i.e.,
\begin{eqnarray}
E_{\rm form,m}^{{\rm Fe}_3{\rm Al}}(V)=\;\;&&E^{{\rm Fe}_3{\rm Al}}_{\rm tot,m}(V)\nonumber\\
&&-\frac{3}{4}\,E_{\rm tot,m}^{\rm bcc-Fe}(V_{\rm eq})
-\frac{1}{4}E_{\rm tot,m}^{\rm fcc-Al}(V_{\rm eq})\quad,\label{eq:eform}
\end{eqnarray}
where $V_{\rm eq}$ marks the respective equilibrium volume of the elemental phase. The 
additional common index $m$ refers to the fact that each energy is given for the same
magnetic state, e.g. nm, pm or fm. Thus {\sl explicit} magnetic-formation/ordering terms 
do not enter our definition of $E_{\rm form}$. In that respect, the data shown in 
Fig.~\ref{fig2:toten} is based on nm-GGA and pm-DFT+DMFT calculations.
Both numerical schemes designate the D0$_3$ structure correctly as the stable one, with 
however two obvious differences. First, while in the many-body scheme the equilibrium volume
is well reproduced, GGA yields a value too small by about 10\%. Second, the energy difference 
$\Delta E_{D0_3}^{L1_2}$ between both structural types is about eight times larger within
DFT+DMFT. Furthermore the bulk modulus $B$ is smaller in the latter scheme.

It was indeed shown in Ref.~\onlinecite{lec02} that the first-principles description of the
electronic structure and the phase stability of Fe$_3$Al is delicate. 
Upon ferromagnetic order, the L1$_2$ phase is by mistake stabilized in GGA(PBE).
In this regard, a detailed data comparison is provided in Tab.~\ref{tab:d03}.
While nm-GGA yields the correct qualitative structural hierachy, the detailed structural data
\begin{table}[t]
\begin{ruledtabular}
\begin{tabular}{l|lccc}
               & $E_{\rm form}$   & $B$      & $a$       & stable \\ \hline
nm-GGA         &  $-394$          & 218      & 5.331     &  $\checkmark$ \\
fm-GGA         &  $-202$          & 151      & 5.465     &  $\lightning$ \\
pm-DFT+DMFT    &  $-325$          & 143      & 5.480     &  $\checkmark$  \\
experiment     &  $-320\pm 20^a$  & 144$^b$  & 5.473$^c$ &  $\checkmark$   \\
\end{tabular}
\end{ruledtabular}
\hfill$^a$\,Ref.~\onlinecite{des87}\;,\; $^b$\,Ref.~\onlinecite{sim71}\;,\;
$^c$\,Ref.~\onlinecite{pea58}\hspace*{4cm}
\caption{Comparison of structural data for D0$_3$-Fe$_3$Al. Formation energy
$E_{\rm form}$ (in meV/atom), bulk modulus $B$ (in GPa). lattice constant (in a.u.)
and stability against the L1$_2$ structure. The (nm,fm,pm) formation energies
are computed using the corresponding (nm,fm,pm) total energy of bcc-Fe 
(cf. eq. (\ref{eq:eform}).}
\label{tab:d03}
\end{table}
\begin{table}[b]
\begin{ruledtabular}
\begin{tabular}{l|cc}
               &  $m_{\rm Fe1}$      & $m_{\rm Fe2}$    \\ \hline
fm-GGA         &  2.45               & 2.12      \\
fm-DFT+DMFT    &  2.17               & 1.48      \\
experiment     &  2.18$^d$, 2.12$^e$ & 1.50$^d$, 1.46$^e$  \\
\end{tabular}
\end{ruledtabular}
\hfill$^d$\,Ref.~\onlinecite{pic61}\;,\; $^e$\,Ref.~\onlinecite{sat95}
\caption{Comparison of the Fe magnetic moments in ferromagnetic D0$_3$-Fe$_3$Al. 
(in $\mu_{\rm B}^{\hfill}$).}
\label{tab:fm}
\end{table}
are off the experimental values. On the good side, introducing ferromagnetism 
on the GGA level brings lattice constant and bulk modulus close to experiment. However it 
not only misleadingly stabilizes the L1$_2$ structure,~\cite{lec02} but now strongly 
underestimates the formation energy. This major difference to the experimental
$E_{\rm form}^{{\rm Fe}_3{\rm Al}}$ appears not to be linked solely to the GGA 
functional, but due to a general insufficient Kohn-Sham description of the magnetic energy 
in Fe-Al. Magnetism has been shown to be important for the D0$_3$ alloy ordering in 
that system.~\cite{kud14} Also in the LDA-based work by Watson and Weinert,~\cite{wat98} 
a value $E_{\rm form,fm}^{{\rm Fe}_3{\rm Al}}=-230\,$ meV/atom
was obtained for spin-polarized D0$_3$-Fe$_3$Al. From the computation of the 
formation energy of various Fe compounds, the authors there concluded that introducing
spin polarization in the Kohn-Sham exchange-correlation functional underestimates the 
magnetic energy for such alloys.

For comparison, we computed also the formation energy of fm-D0$_3$
within DFT+DMFT. The corresponding value $E_{\rm form,fm}^{{\rm Fe}_3{\rm Al}}=-315\,$
meV/atom differs only little from the pm value. Thus the magnetic-ordering energy does 
not strongly influence the D0$_3$ ordering, when assuming coherent magnetic states. 
Of course, the {\sl explicit} magnetic-ordering energy 
$E_{\rm form,fm}^{{\rm Fe}_3{\rm Al}}-E_{\rm form,pm}^{{\rm Fe}_3{\rm Al}}=-170\,$
meV/atom is still sizable. Concerning the competition between chemical orderings
with pm- {\sl or} fm-order in the Fe-Al phase diagram, this latter energy is surely relevant.
A detailed statistical-mechanics study of this problem is however beyond the scope
of the present work. For completeness, we provide in Tab.~\ref{tab:fm} the magnetic
moments in fm-D0$_3$. While GGA overestimates the local-Fe moments, DFT+DMFT once more
brings the data in line with experimental findings.

The results of the DFT+DMFT scheme are overall in very good agreement with the available
experimental data. Note again that in order to evaluate the formation energy, the bcc-Fe 
problem was of course also treated in DFT+DMFT, respectively with the same magnetic state
$m$ and with identical local Coulomb interactions. Compared to nm-GGA, the 
less negative $E_{\rm form}$ of 
the D0$_3$ structure, in better agreement with experiment, matches the reduced bonding 
strength revealed from the correlated charge densities (cf. Fig.\ref{fig5:charge}). 
For the case of L1$_2$-Fe$_3$Al, correlations not only
render it much more energetically unfavorable compared to D0$_3$. Its formation energy
becomes even positive, marking the compound unstable. This may be explained by the
discussed correlation-induced weakening of the $xz/yz$ states with significant 
bonding-antibonding nature, compared to the strengthening of the $xy$ and $z^2$ states.
Thus many-body effects beyond conventional DFT do not merely lead to some quantitative 
changes, but display a {\sl qualitative} effect on the energetics in the Fe-Al system. 

The general improvement in the theoretical description of D0$_3$-Fe$_3$Al underlines 
not only the importance of electronic correlations, but renders it clear that a faithful 
description of the paramagnetic state is sufficient to account for the phase-relevant 
characterization.

\subsection{Fe$_2$VAl}
In the final section, we discuss the electronic structure of the Heusler 
L2$_1$-Fe$_2$VAl compound that emerges from D0$_3$-Fe$_3$Al by replacing the Fe1 
sublattice with V atoms.

\begin{figure}[b]
(a)\hspace*{-0.4cm}\includegraphics*[width=8.65cm]{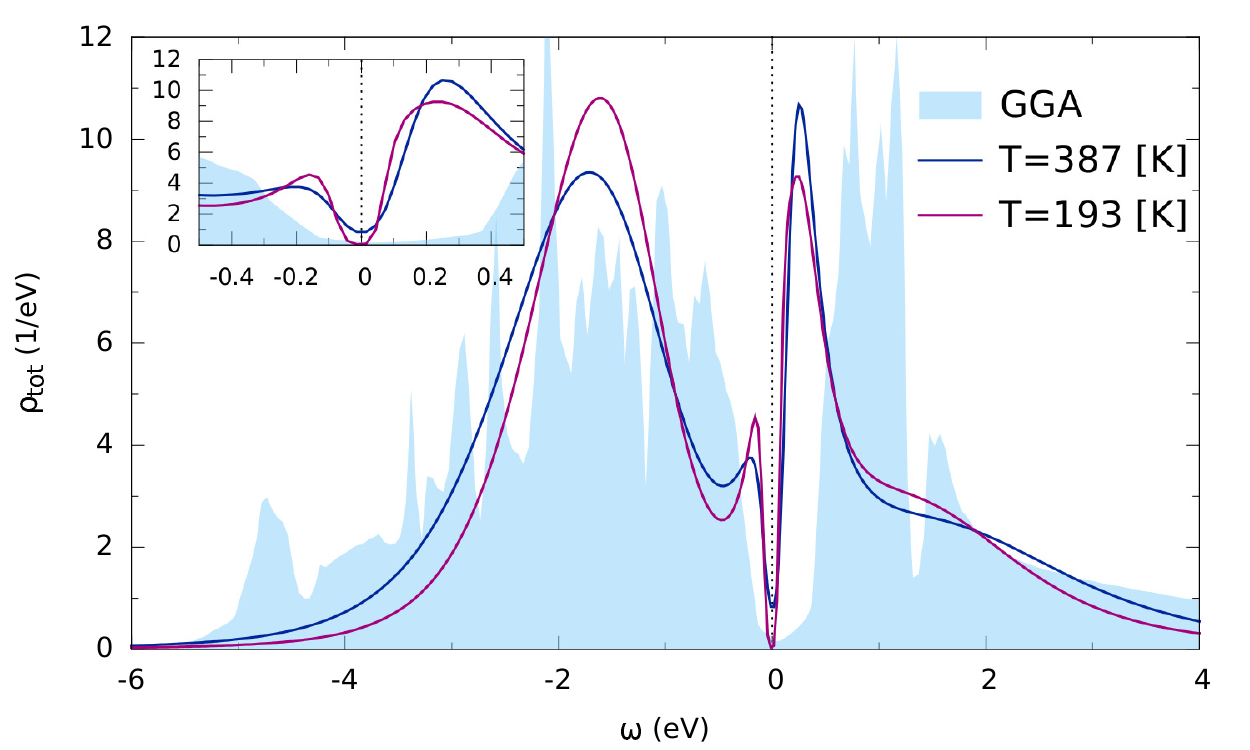}\\
(b)\hspace*{-0.4cm}\includegraphics*[width=8.65cm]{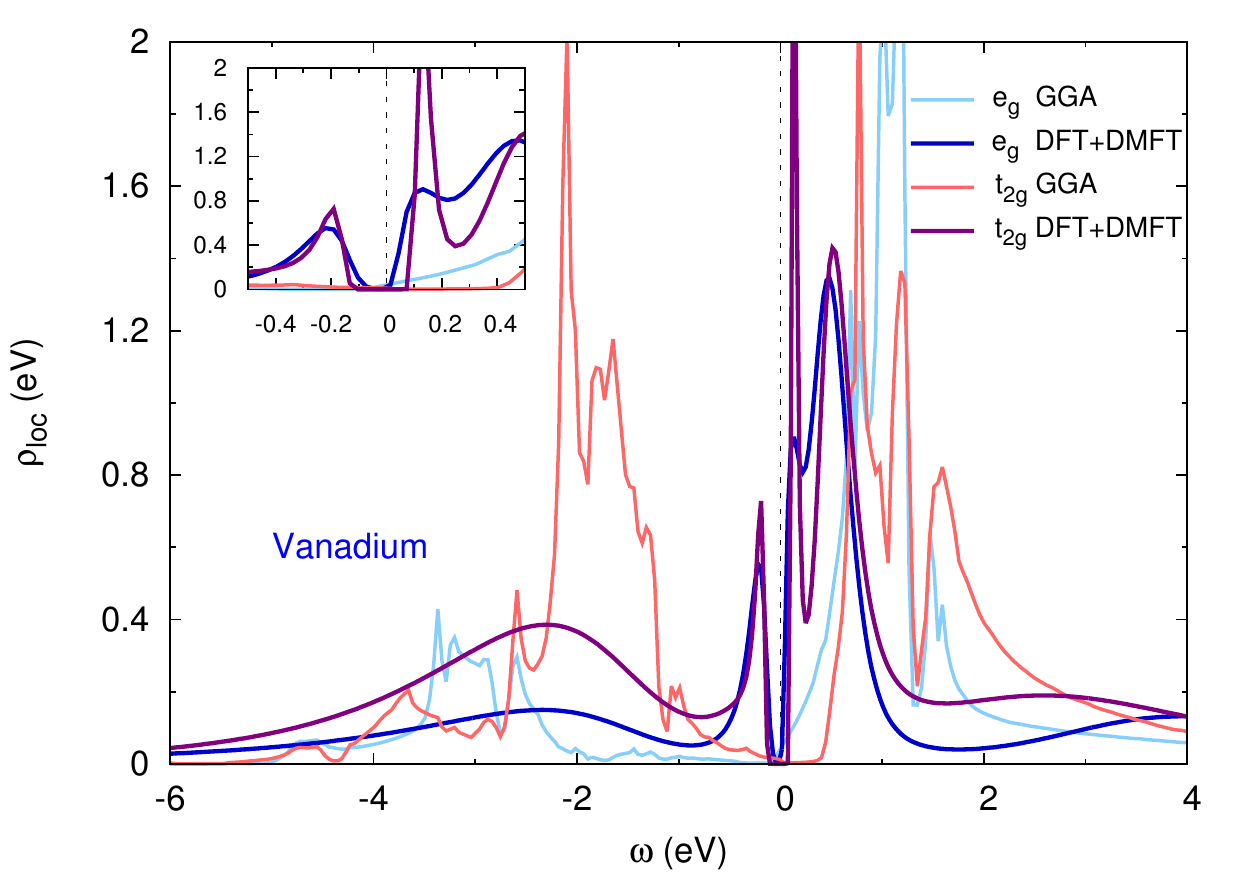}\\
(c)\hspace*{-0.4cm}\includegraphics*[width=8.65cm]{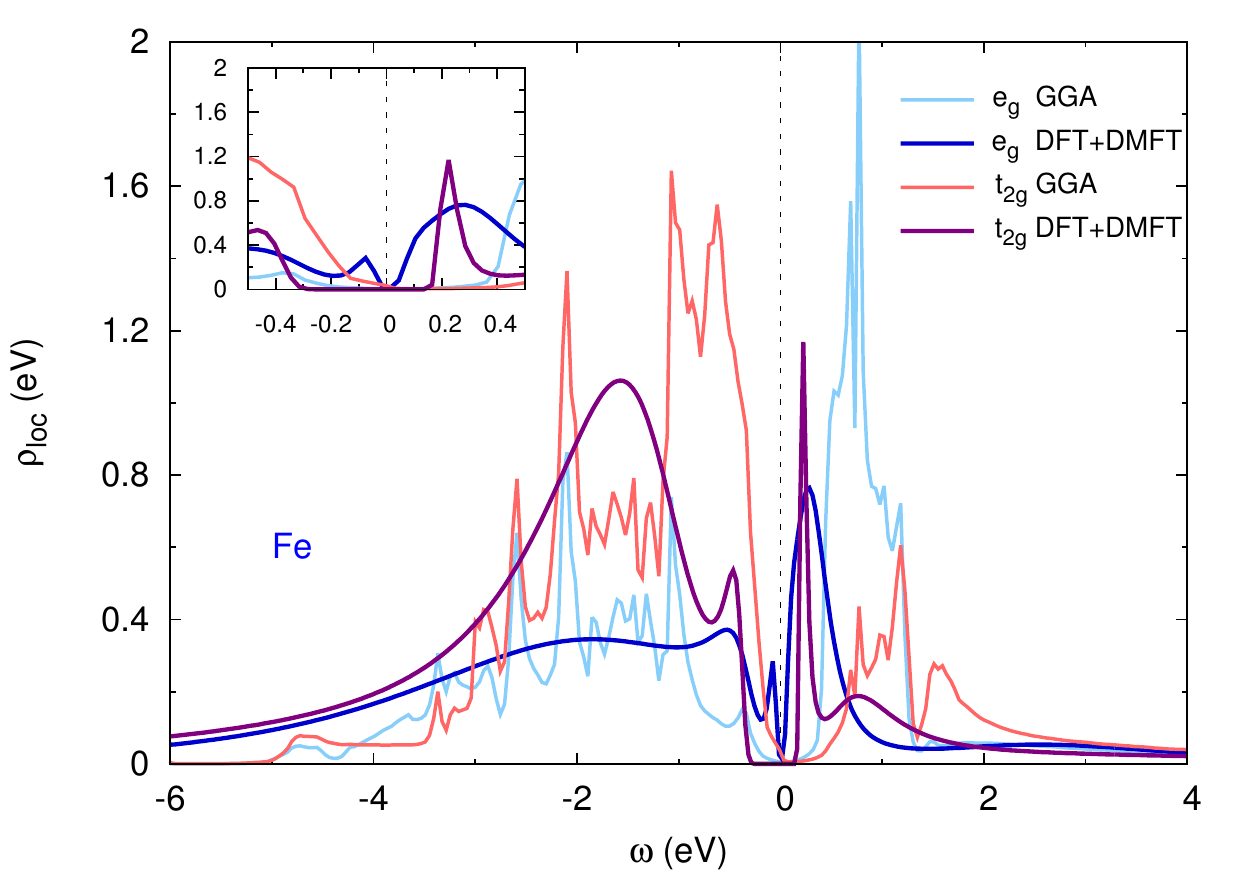}\\
\caption{(color online) Spectral function of L2$_1$-Fe$_2$VAl.
Top: total with DFT+DMFT for two different temperatures, 
middle: V local, bottom: Fe local.}\label{fig6:l21spec}
\end{figure}

Figure~\ref{fig6:l21spec} shows the total and local spectral properties, again by
comparing GGA and DFT+DMFT. As in the earlier DFT-based studies~\cite{sin98,weh98} we
find a semimetallic solution in the former Kohn-Sham calculation. A dichotomy is seen by 
inspecting the transition-metal electron-state characters on the local level. 
Below the Fermi level the low-energy region is dominated by Fe($t_{2g}$) states, while
above $\varepsilon_{\rm F}$ there are dominantly V($e_g$) states. As expected because of
replacing the Fe1 ions, the V site has a more pronounced orbital differentiation. Yet
due to the different vanadium valence, the GGA filling is of course only a bit more 
than half the size of the Fe site.

Note that within DFT+DMFT we utilize the same $U$ and $J_{\rm H}$ on the Fe and
V sites. This choice can be motivated based on the strong inter-site hybridizations
in the given intermetallic system, leading to a coherent screening that minimizes
substantial site differences in the local Coulomb interactions. With correlations, 
a clear gap structure emerges in the low-energy region, which is only
fully realized at lower temperature. A pseudogap signature is obtained at a higher
$T=387$\,K. It is notable that spectral weight is shifted towards the low-energy
region to form sharp gap edges. Thus the gap formation is not of obvious single-particle
character and has similarities with e.g. the (pseudo)gap structure in cuprates. Therefore
the insulating state is not of conventional band-insulating semiconductor type. Measuring 
the charge-gap from the mids of the gap-edge structure, a size $\Delta_{\rm g}\sim 0.15$\,eV 
is read off at $T=193$\,K. This is in excellent agreement with experimental values for a 
charge gap in Fe$_2$VAl.~\cite{lue98}

Both transition-metal elements contribute to the intricate gap formation, but the V ion
seems to play a more dominant role due to the larger spectral-function enhancement at
the gap edges. Moreover the low-energy spectra with correlations displays are more
balanced $e_g, t_{2g}$ contribution compared to GGA. This is in line with a nearly
orbital-independent local self-energy on the V sites. Therefrom, the correlation strength
is enhanced on the latter sites in comparison to the Fe sites, yet the vanadium-based 
QP weight $Z\sim 0.7$ is again moderate. Needless to say that Fe$_2$VAl is of course no
Mott insulator. Still electronic correlations beyond conventional DFT are at the 
origin of the gap formation and -opening. Interesting in this context are the different
electron fillings of V and Fe (see Tab.~\ref{tab3:occ2}).
While the Fe ion non-surprisingly shows a very similar filling characteristic as the Fe2
ion in D0$_3$-Fe$_3$Al, the V ion already surely differs in the number of valence 
electrons. With an effective filling close to four electrons, the V site lies one hole
below half filling, i.e. in a designated Hund's metal regime.~\cite{wer08,hau09,med11} 
The orbital-resolved V occupations align somewhat in DFT+DMFT, however it seems that 
the overall correlation strength due to the given sizes of bandwidth and local Coulomb 
interactions is yet too weak to drive very strong Hund's physics. But unconventional 
spin fluctuations could nonetheless play a relevant role in the enhanced experimental 
specific heat.~\cite{lue99}

\begin{table}[t]
\begin{ruledtabular}
\begin{tabular}{l|lcccccr}
             &          & $e_{g}$   &       & $t_{2g}$     &   &      &  total\\ 
             &          & $x^2-y^2$ & $z^2$ & $xz$  & $yz$ & $xy$ &  \\ \hline
             &          & 0.45      & 0.45  & 0.97  & 0.97 & 0.97 &\hspace*{0.5cm}3.81\\ 
             & \bw{V}   & 0.70      & 0.70  & 0.90  & 0.90 & 0.90 &\hspace*{0.5cm}4.10\\[0.1cm] 
\bw{L2$_1$}  &          & 1.09      & 1.09  & 1.65  & 1.65 & 1.65 &\hspace*{0.5cm}7.13\\ 
             & \bw{Fe}  & 1.45      & 1.45  & 1.72  & 1.72 & 1.72 &\hspace*{0.5cm}8.06\\
\end{tabular}
\end{ruledtabular}
\caption{Projected-local-orbital occupations in Fe$_2$VAl for $T=387\,$K. First lines are
GGA, second lines DFT+DMFT results, respectively.}\label{tab3:occ2}
\end{table}

\section{Conclusions}
Recently, there have been various investigations that employ realistic DMFT approaches 
beyond Kohn-Sham DFT(+U) to elemental iron and its alloy with aluminum. Studies on 
phase stabilities in the Fe phase diagram,~\cite{leo11,pou13} on the $\alpha$-Fe phonon 
spectrum,~\cite{leo12} on vacancy formation in $\alpha$-Fe,~\cite{del16} and on
the magnetism in B2-FeAl~\cite{pet03,gal15} were performed. The present work adds 
the DFT+DMFT examination of the Fe$_3$Al and Fe$_2$VAl correlated electronic structure. 

We show that albeit both compounds do not fall in the standard category of strongly 
correlated systems, more subtle many-body effects are still relevant for a correct 
description. The energetics and phase stability of Fe$_3$Al builds upon such effects, 
by providing an improved value for the formation energy with a clear energy separation 
to the L1$_2$ structure. Note that the charge self-consistent version of the DFT+DMFT
framework is important to elucidate such physics. Thereby not only local changes on
the explicitly correlated lattice sites matter, but the overall charge redistribution 
including also interstitial and ligand region are crucial. On general grounds for cubic 
intermetallics, the open bcc lattice seems more adequate for correlated (Fe-based) 
compounds. For systems on the close-packed fcc lattice with sizeable local Coulomb 
interactions, the local correlations become comparatively too strong, weakening important
bonding properties. Fcc-based compounds like e.g. Ni$_3$Al and Cu$_3$Au either do not 
display substantial {\sl local} correlation physics or are well described in standard DFT.
We want to note that the issue of chemical disorder is surely relevant concerning
the phase stabilities close to the Fe$_3$Al composition of the Fe-Al phase 
diagram.~\cite{sun09} Treating such additional degrees of freedom together with the 
here encountered correlations beyond DFT is a formidable task which has to be faced in the 
future for a detailed thermodynamic understanding of Fe-Al.

The Fe$_2$VAl compound manifests an intriguing twist to traditional intermetallics. In
the sense that the material is derived from the well-known Fe$_3$Al metal but
displays an intricate gap opening reminiscent of (pseudo)gap physics observed in correlated
oxides. The DFT+DMFT gap size and its sensitivity to temperature are in excellent agreement
with experimental results for this compound. Since also Hund's physics may play a
role on the vanadium site, this example shows how easily traditional materials physics 
may be confronted with challenging mechanisms from strongly correlated matter. 
In a future work, addressing the thermoelectric properties of Fe$_2$VAl on the basis of 
the here established results would be highly interesting.

\begin{acknowledgments}
We thank D. Grieger and M. Obermeyer for helpful discussions.
This research was supported by the DFG-FOR1346.
Computations were performed at the University of Hamburg 
and at the North-German Supercomputing Alliance (HLRN) under Grant No. hhp00026.
\end{acknowledgments}

\bibliography{bibextra}

\end{document}